\documentclass[a4paper,aps,pre,superscriptaddress,floatfix,nofootinbib,longbibliography,notitlepage,onecolumn]{revtex4-1}
\usepackage{bbm}
\usepackage{bbold}
\usepackage{bbm}
\usepackage[pdftex]{graphicx}
\usepackage{latexsym,amsmath,verbatim,amssymb,txfonts}
\usepackage{lipsum}
\usepackage{color}
\usepackage{rotating}
\usepackage{verbatim}
\usepackage{multirow}
\usepackage[english]{babel}
\usepackage{comment}
\usepackage{bm}
\usepackage{amsmath}
\usepackage{amsfonts}
\usepackage{amssymb}
\usepackage{color}
\usepackage{braket}
\usepackage{blkarray, bigstrut}

\definecolor{internationalkleinblue}{rgb}{0.0, 0.18, 0.65}
\definecolor{brickred}{rgb}{0.8, 0.25, 0.33}
\definecolor{amber}{rgb}{1.0, 0.75, 0.0}
\definecolor{applegreen}{rgb}{0.55, 0.71, 0.0}

\newcommand{\B}[1]{\if#1\relax\bm{#1}\else\mathbf{#1}\fi} 

\begin{document}
\title{Pinning control of chimera states in systems with higher-order interactions}

\author{Riccardo Muolo}
\email{corresponding author: muolo.r.aa@m.titech.ac.jp}
\affiliation{Department of Systems and Control Engineering, Institute of Science Tokyo (former Tokyo Institute of Technology), Tokyo 152-8552, Japan}

\author{Lucia Valentina Gambuzza} 
\affiliation{Department of Electrical, Electronics and Computer Science Engineering, University of Catania, 95125 Catania, Italy}

\author{Hiroya Nakao}
\affiliation{Department of Systems and Control Engineering, Institute of Science Tokyo (former Tokyo Institute of Technology), Tokyo 152-8552, Japan}
\affiliation{Research Center for Autonomous Systems Materialogy, Institute of Innovative Research, Institute of Science Tokyo (former Tokyo Institute of Technology), Kanagawa 226-8501, Japan}

\author{Mattia Frasca }
\affiliation{Department of Electrical, Electronics and Computer Science Engineering, University of Catania, 95125 Catania, Italy}

\date{\today}

\begin{abstract}

Understanding and controlling the mechanisms behind synchronization phenomena is of paramount importance in nonlinear science. In particular, the emergence of chimera states, patterns in which order and disorder coexist simultaneously, continues to puzzle scholars, due to its elusive nature. Recently, it has been shown that higher-order {(many-body)} interactions greatly enhance the presence of chimera states, which are {easier to be found and more persistent.} In this work, we show that the higher-order framework is fertile not only for the emergence of chimera states, but also for its control. Via pinning control, a technique consisting in applying a forcing to a subset of the nodes, we are able to trigger the emergence of chimera states with only a small fraction of controlled nodes, at striking contrast with the case without higher-order interactions. We show that our setting is robust for different higher-order topologies and types of pinning control and, finally, we give a heuristic interpretation of the results via phase reduction theory. Our numerical and theoretical results provide further understanding on how higher-order interactions shape {collective behaviors in} nonlinear dynamics.   


\end{abstract}

\maketitle

\section{Introduction}

Understanding the mechanisms underlying self-organization phenomena on networks is a paramount task in the study of complex systems, which is complemented by the development of efficient methods to control such dynamics \cite{d2023controlling}. This is particularly relevant in the framework of synchronization dynamics, where, depending on the applications, it is fundamental to achieve a synchronized state, e.g., in power grids \cite{sajadi2022synchronization}, or to break it into an asynchronous one, e.g., in neuroscience \cite{asllani2018minimally}, where synchronization is often associated to pathological states. The network framework remains still relevant in the modeling of complex systems, nonetheless, over the past years scholars have started considering more complex structures such as hypergraphs and simplicial complexes \cite{battiston2020networks,battiston2021physics,bianconi2021higher,boccaletti2023structure,bick2023higher,millan2025topology}. This is because networks do not capture interactions beyond the pairwise setting, i.e., two-by-two, while many systems have shown evidence of higher-order, i.e., group, interactions \cite{battiston2020networks,battiston2021physics}. Examples come from, but are not limited to, neuroscience \cite{rosen1989,wang2013,petri2014homological,sizemore2018cliques}, ecology \cite{grilli_allesina,iacopini2024not} and social behaviors \cite{centola2018experimental}. Higher-order interactions have been proven to greatly affect the collective behavior, for instance, in random walks \cite{carletti2020random,schaub2020random}, synchronization dynamics \cite{tanaka2011multistable,skardal2019higher,millan2020explosive}, contagion \cite{iacopini2019simplicial} and pattern formation \cite{carletti2020dynamical,muolo2024turing}, to name just a few. Given the ubiquity of group interactions \cite{battiston2020networks,battiston2021physics,bianconi2021higher,boccaletti2023structure,bick2023higher}, it is important to understand how to control the dynamics in such systems. While significant progress has been made in the control of networks \cite{liu2011controllability,liu2016control}, the investigation into the control of systems with higher-order interactions has only recently begun \cite{chen2021controllability,de2022pinning,de2023pinning,xia2024pinning,moriame2025hamiltonian}. 

The focus of this work is an intriguing type of synchronization pattern called chimera state, which consists of the coexistence of coherent and incoherent domains of oscillations. Coexistence of coherence and incoherence was first observed by Kaneko for globally coupled chaotic maps \cite{kaneko} and was then found in several numerical settings with global \cite{hakim1992dynamics,nakagawa1993collective,chabanol1997collective} and nonlocal \cite{kuramoto1995scaling,kuramoto1996origin,kuramoto1997power,kuramoto1998multiaffine,kuramoto2000multi} coupling schemes. Despite all the previous research on the subject, the work that historically is considered to be the first to characterize the emergence of chimera states is the well-known paper by Kuramoto and Battogtokh \cite{kuramoto2002coexistence}, made popular by a successive work of Abrams and Strogatz, who, with a creative intuition, compared the coexistence of different dynamical state to the chimera, a mythological creature in which parts of different animals coexisted \cite{abrams2004chimera}. Besides the pure theoretical relevance of such an astonishing phenomenon, a great part of the interest has been generated by the existence of analogous patterns in real systems: for instance, in Josephson junctions \cite{cerdeira_prl} and electronic circuits \cite{vale_chimeras,gambuzza2020experimental}, laser \cite{hagerstrom2012experimental}, mechanical \cite{martens2013chimera} and nano-electromechanical systems \cite{matheny2019exotic}, to name a few. Particular attention has been devoted to neuroscience \cite{chimera_neuro,majhi2019chimera}, specifically to unihemispheric sleeping patterns in animals \cite{rattenborg2000behavioral}. Except for some particular configurations in which robust chimera patterns are induced by the network structure \cite{bram_malb_chim,muolo_2023_chimera,asllani2025pattern}, { and for which a rigorous stability analysis can be carried out,} in both numerical and experimental settings chimera states are often elusive and characterized by a rather short lifetime{. Note that, if the coupling is global (all-to-all), chimeras may not always be transient in the thermodynamic limit.} Hence, there is a vast literature on networked systems, consisting in looking for different settings (e.g., parameter ranges, network topologies, coupling configurations, etc.) making such patterns easier to find and with a longer lifetime. Moreover, after the first definition by Kuramoto and Battogtokh \cite{kuramoto2002coexistence}, several kinds of chimera states have been defined, e.g., amplitude chimeras \cite{zakharova2016amplitude} or phase chimeras \cite{zajdela_abrams}. We will not thoroughly discuss such studies, inviting the interested reader to consult a book \cite{zakharova2020chimera} and a review \cite{parastesh2021chimeras} on the subject. In the context of higher-order interactions, chimera states have been proven to be enhanced in some pioneering works considering both pairwise and higher-order interactions \cite{kundu2022higher,ghosh_chimera2,bick_nonlocal1}. This claim was further corroborated in \cite{muolo2024phase} for systems with pure higher-order interactions, where the emergence of chimera states on higher-order topologies {was} compared with the absence of such patterns when the interactions are pairwise. 

{Several techniques have been developed to control the emergence of chimera states on networks, including, but not limited to, gradient dynamics \cite{bick2015controlling}, feedback control \cite{omelchenko2018optimal,omelchenko2019control}, delay \cite{semenov2016deterministic,ghosh2018engineering,ghosh2019taming}, pace makers \cite{ruzzene2020remote}, and time varying networks \cite{majhi2022oscillation}, to name a few.} In this work, we consider the setting studied in \cite{muolo2024phase} and implement a control to further trigger the emergence of chimera states. Our control approach will rely on the so called \textit{pinning control}, a technique used to drive networks onto a desired dynamical state by using a control input applied to a small subset of nodes \cite{grigoriev1997pinning,sorrentino2007controllability}. Such technique has been successfully used in the framework of opinion dynamics \cite{iudice2022bounded,ancona2023influencing}, epidemics \cite{du2015selective,yang2019feedback}, pattern formation \cite{buscarino2019turing} and synchronization dynamics \cite{porfiri2008criteria,yu2013synchronization,gambuzza2016pinning}, to name a few. The latter includes the control of chimera states with pairwise interactions, which we hereby extend to the higher-order framework. Indeed, in \cite{gambuzza2016pinning} it was shown that it is possible to trigger the emergence of chimera states via pinning. Nonetheless, to achieve such task, at least half of the network nodes need to be controlled. In what follows, after introducing the setting in Sec. \ref{sec:model}, we show that higher-order interactions considerably facilitate the work of the controllers and chimera states can be obtained by controlling only a small fraction of the nodes. Such results are shown in Sec. \ref{sec:numerics} for two different kinds of pinning approaches, that we named additive pinning and parametric pinning. Moreover, we show that, rather than the number of nodes, what matters is the size of the hyperedges, i.e., the group of nodes interacting with each other. Then, before the discussion of some potential future directions, we give a heuristic interpretation of the results based on {the theory of phase reduction} \cite{nakao2016phase} in Sec. \ref{sec:heuristic}.

\section{The model and the setting}\label{sec:model}

In this Section, we introduce the system exhibiting chimera states, which is analogous to that studied in \cite{muolo2024phase}. We consider coupled Stuart-Landau oscillators, a paradigmatic model for the study of synchronization dynamics, given that it is the normal form of the supercritical Hopf-Andronov bifurcation \cite{nakao2014complex}. The coupling takes place through pure higher-order interactions, namely, by mean of a higher-order topology called \textit{nonlocal hyperring}, which is a generalization of the nonlocal pairwise coupling \cite{muolo2024phase}. The type of chimera state that we will hereby consider is that of \textit{phase chimeras}, states have been first observed by Zajdela and Abrams \cite{zajdela_abrams}, which consist in oscillation patterns where the amplitude and the frequency of each oscillator are the same, but the phases exhibit a chimera behavior, i.e., a part of the oscillators have the same phase, while the other phases are distributed along the unit circle. The peculiarity of such pattern is that, once obtained, it does not vary, because the frequency is the same for all the oscillators. Hence, we would observe the same exact pattern after each period. For this reason, we find the description given by Zajdela and Abrams, "frozen patterns of disorder", perfectly fitting. The reader can find a thorough characterization of these patterns in the Refs. \cite{zajdela_abrams,muolo2024phase}{, which are not completely "frozen" in the latter, the coupling not being all-to-all.} On a side, let us note that multitailed phase chimeras have only been found in the pairwise setting \cite{zajdela_abrams}, but not yet in the higher-order one. In what follows, every chimera state discussed and shown in the figures will be a phase chimera. For sake of simplicity, we will refrain from using the word "phase" and will call them simply "chimeras".

\subsection{Stuart-Landau oscillators coupled via nonlocal hyperrings}

We consider a system made of $N$ interacting Stuart-Landau units. In the absence of any interaction, each unit $i$ of the system is described by the following equations
\begin{equation}
   \begin{cases}\label{eq:SL}
 \dot{x}_i = \alpha  x_i - \omega y_i  -\left(x_{i}^2  + y_{i}^2 \right)x_{i}=f(x_i,y_i), \\\\
 \dot{y}_i = \omega x_i + \alpha y_i - \left(x_{i}^2  + y_{i}^2 \right)y_{i}=g(x_i,y_i)\, , \end{cases}
\end{equation}
where $\alpha$ is a bifurcation parameter and $\omega$ is the frequency of the oscillators. Let us stress that the units are homogeneous, meaning that the parameters $\alpha$ and $\omega$ are the same for each and every system. Each isolated system exhibits a stable limit cycle for $\alpha>0$, which is the case we will consider throughout this study. {Note that the above formulation of the Stuart-Landau equations comes from the Complex Ginzburg-Landau Equation (CGLE), from which real and imaginary parts are taken, which are variables $x$ and $y$, respectively \cite{nakao2014complex}.}

\begin{figure}
\includegraphics[scale=0.14]{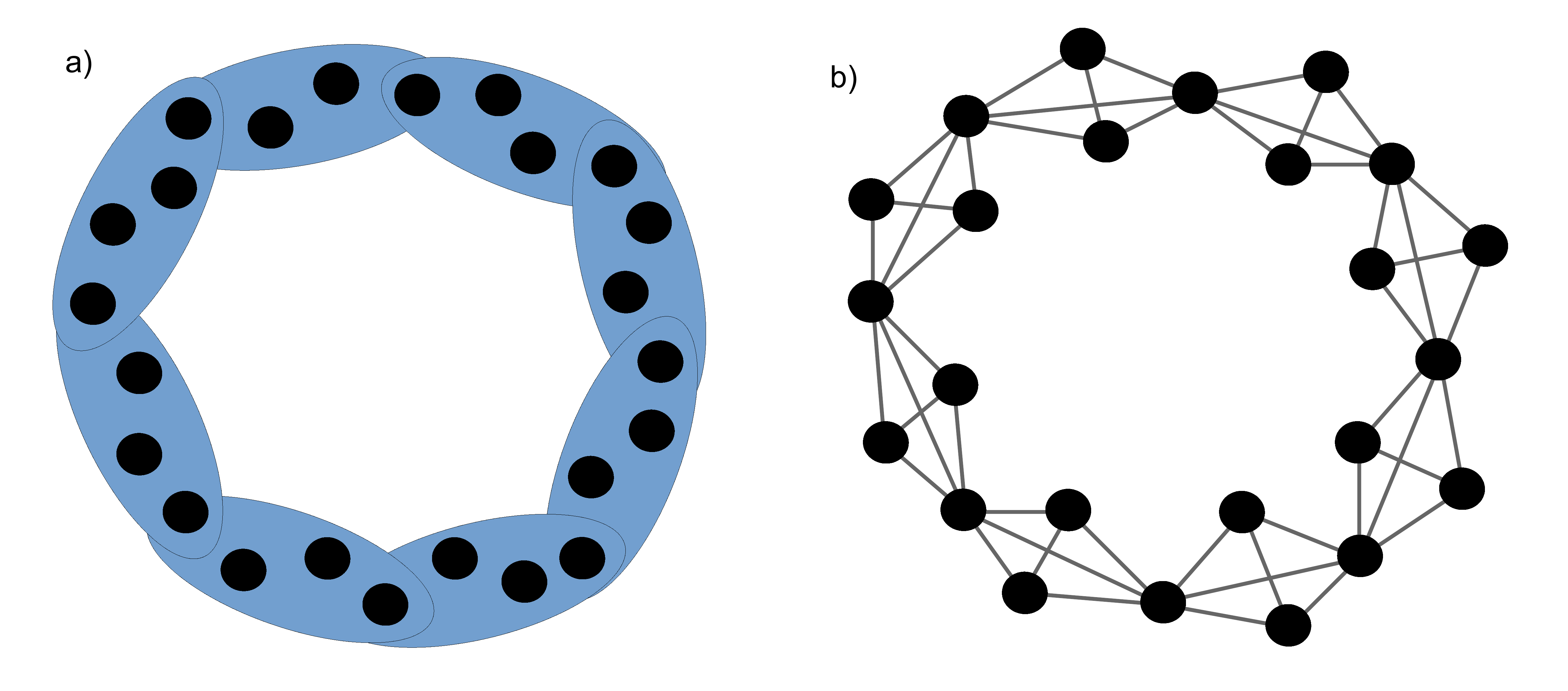}
\caption{In panel a), a $3$-hyperring of $24$ nodes, together with its corresponding clique-projected network, depicted in panel b). We will call the nodes which are part of two hyperedges (resp. cliques) \textit{junction nodes}, while all the others will be called \textit{non-junction nodes}.} \label{fig:3hyp_vs_clique}
\end{figure}

To model the higher-order interactions we use a generalization of the links (or edges) called \textit{hyperedges}, whose structure can be encoded by using \textit{adjacency tensors}, a generalization of the adjacency matrices for networks \cite{battiston2020networks}. From the literature dealing with simplicial complexes \cite{bianconi2021higher}, we adopt the convention that a hyperedge of $(d+1)$ nodes (i.e, encoding a $(d+1)$-body interaction) is called a $d$-hyperedge. As an example, let us consider the $3$-rd order\footnote{Let us note that in tensor algebra the order (or rank) of a tensor is given by its indices, e.g., a scalar is a $0$-rank tensor, a vector is a $1$-rank tensor, a matrix is a $2$-rank tensor, etc. Hence, $A^{(3)}=\{A_{ijkl}^{(3)}\}$ would be a $4$-rank tensor and the adjacency matrix, given that it is a matrix, a $2$-rank tensor. However, here we follow that notation that is most commonly used in the literature on higher-order interactions.} adjacency tensor (i.e., encoding $4$-body interactions) $A^{(3)}=\{A_{ijkl}^{(3)}\}$. We have that $A_{ijkl}^{(3)}=1$ if nodes $i,j,k,l$ are part of the same hyperedge and $0$ otherwise. This is the analogous of the adjacency matrix for networks, which is, indeed, a $1$-st order adjacency tensor. The chosen higher-order topology is that of \textit{nonlocal $d$-hyperrings}, a pure higher-order structure introduced in \cite{muolo2024phase} and consisting of hyperedges of size $d+1$ ($d$ is the order of the interaction) attached through one node and set in a circular structure. In Fig. \ref{fig:3hyp_vs_clique}, we depict a $3$-hyperring, in panel a), together with its pairwise counterpart, namely, the \textit{clique-projected network (cpn)} obtained by projecting each hyperedge into a clique having the same size, in panel b). Observe that a hyperring is a uniform hypergraph, the hyperedges all having the same size. 

Following the analyses carried out in previous works \cite{gambuzza2020experimental,muolo2024phase}, we assume the coupling to involve only the first state variable of each oscillator, i.e., $x$. Let us stress that other coupling configurations can be considered, as discussed in Appendix \ref{sec:appA}. With the above assumption, the equations for systems \eqref{eq:SL} coupled via a generic $d$-hyperring read

\begin{equation}
\begin{cases}
\displaystyle \dot{x}_i =f(x_i,y_i) + \varepsilon \sum_{j_1,...,j_d=1}^N A_{ij_1...j_d}^{(d)}\Big(h^{(d)}(x_{j_1},...,x_{j_d}) - h^{(d)}(x_i,...,x_i) \Big),\\ \\
\displaystyle \dot{y}_i =g(x_i,y_i)+ \varepsilon \sum_{j_1,...,j_d=1}^N A_{ij_1...j_d}^{(d)}\Big(h^{(d)}(x_{j_1},...,x_{j_d})- h^{(d)}(x_i,...,x_i) \Big),
\end{cases}
\label{eq:SL_HO}
\end{equation}
where $\varepsilon>0$ is the coupling strength\footnote{Note that such configuration involves only interactions of order $d$, i.e., $(d+1)$-body, hence it is not necessary to denote the coupling strength with $\varepsilon_d$.}. Because we consider identical oscillators, together with the fact that the coupling is {diffusive-like}, i.e., it vanishes when all oscillators are in the same state, system \eqref{eq:SL_HO} admits a fully synchronous solution. Such a coupling is a special type of non-invasive interaction \cite{gambuzza2021stability}. Moreover, we will consider coupling functions such that the higher-order interaction cannot be decomposed into pairwise ones\footnote{It was shown by Neuhäuser et al. that the higher-order coupling functions need to be nonlinear, otherwise the many-body interaction can be decomposed into pairwise ones \cite{neuhauser2020multibody}. Successively, it was further shown that such assumption may not be enough and, if the nonlinear functions $h$ are the sum of nonlinear terms which separately account for each unit, e.g., $h^{(d)}(x_{j_1},...,x_{j_d})=h(x_{j_1})+...+h(x_{j_d})$, they can still be decomposed into pairwise ones \cite{muolo2023turing}.}. Eq. \eqref{eq:SL_HO} can be rewritten in compact form for each unit $i$ as \begin{equation}\label{eq:compact}
    \dot{\vec{X}}_i=\vec{F}(\vec{X}_i)+\mathbb{D}\sum_{j_1,...,j_d=1}^N A_{ij_1...j_d}^{(d)}\Big(\vec{H}^{(d)}(\vec{X}_{j_1},...,\vec{X}_{j_d})-\vec{H}^{(d)}(\vec{X}_{i},...,\vec{X}_i)\Big),
\end{equation} where $\vec{X}_i=(x_i,y_i)^\top$, $\vec{F}=(f,g)^\top$, $\vec{H}^{(d)}=(h^{(d)},h^{(d)})^\top$ and $\mathbb{D}=\varepsilon D=\varepsilon\begin{bmatrix}
    1& 0 \\ 1 & 0
\end{bmatrix}$. {As previously stated}, in Appendix \ref{sec:appA} we report additional results for different coupling matrices $D$.\\
To highlight the effects of higher-order interactions, we will perform a comparison between the dynamics on the hyperring and on its respective clique-projected network (cpn), as in \cite{muolo2024phase}. The equations for the dynamics with pairwise interactions are 
\begin{equation}
\begin{cases}
\displaystyle \dot{x}_i =f(x_i,y_i) + \varepsilon \sum_{j=1}^N A_{ij}^{(1)}\Big(h^{cpn}(x_j) - h^{cpn}(x_i) \Big),\\ \\
\displaystyle \dot{y}_i =g(x_i,y_i)+ \varepsilon \sum_{j=1}^N A_{ij}^{(1)}\Big(h^{cpn}(x_j)- h^{cpn}(x_i) \Big),
\end{cases}
\label{eq:SL_clique}
\end{equation} where the coupling functions for the dynamics on the clique-projected network $h^{cpn}$ is determined from its corresponding $h^{(d)}$ (see below).

We will perform pinning control on hyperrings of $4$ different orders, namely, $3$-,$4$-,$5$- and $6$-hyperrings, involving $4$-,$5$-,$6$- and $7$-body interactions, respectively. { For what concerns the functional form of the nonlinear coupling, we will choose odd-degree polynomials. This is because when Stuart-Landau (SL) oscillators are coupled through higher-order couplings of polynomial type with even degree, they behave as if there were no coupling. Indeed, through phase reduction theory, it can be shown that for oscillators with symmetry properties, such as the SL, higher-order couplings of polynomial type with even degree do not have any effect on the phase model at the first order approximation \cite{leon2025theory}.} The coupling functions for each hyperring and its clique-projected network are the following

\begin{equation}
 \label{eq:coupl}
 \begin{cases}
     h^{(3)}(x_{j_1},x_{j_2},x_{j_3})=x_{j_1}x_{j_2}x_{j_3} ~~~~~~~~~~~~~~~~~~~~~~~~~~~~~~~~~~~,~~~h^{cpn}(x_j)=x_j^3 , \\\\
     h^{(4)}(x_{j_1},x_{j_2},x_{j_3},x_{j_4})=x_{j_1}^2x_{j_2}x_{j_3}x_{j_4} ~~~~~~~~~~~~~~~~~~~~~~~~,~~~h^{cpn}(x_j)=x_j^5 , \\\\
     h^{(5)}(x_{j_1},x_{j_2},x_{j_3},x_{j_4},x_{j_5})=x_{j_1}x_{j_2}x_{j_3}x_{j_4}x_{j_5} ~~~~~~~~~~~~~~,~~~h^{cpn}(x_j)=x_j^5 , \\\\
     h^{(6)}(x_{j_1},x_{j_2},x_{j_3},x_{j_4},x_{j_5},x_{j_6})=x_{j_1}^2x_{j_2}x_{j_3}x_{j_4}x_{j_5}x_{j_6} ~~~,~~~h^{cpn}(x_j)=x_j^7 ,
    \end{cases}
\end{equation}
where the adjacency tensor accounts for all the permutations of the indexes. { Note that the pairwise coupling for the systems on the clique-projected networks of the $4$- and $5$-hyperrings have the same functional form. We will come back to this point when discussing the dynamics of the systems with pairwise interactions.} As an example, {let us now explicitly write} the equations for the $4$-body case ($3$-hyperring) and its corresponding clique-projected network. The equations for a $3$-hyperring are the following 

\begin{equation}
\begin{cases}
\displaystyle \dot{x}_i =\alpha  x_i - \omega y_i  -\left(x_{i}^2  + y_{i}^2 \right)x_{i} + \varepsilon \sum_{j_1,j_2,j_3}^N A_{ij_1j_2j_3}^{(3)}\Big(x_{j_1}x_{j_2}x_{j_3} - x_i^3\Big),\\ \\
\displaystyle \dot{y}_i =\omega x_i + \alpha y_i - \left(x_{i}^2  + y_{i}^2 \right)y_{i}+ \varepsilon \sum_{j_1,j_2,j_3}^N A_{ij_1j_2j_3}^{(3)}\Big(x_{j_1}x_{j_2}x_{j_3} - x_i^3 \Big),
\end{cases}
\label{eq:3hyp}
\end{equation}

\noindent while those for the clique-projected network are

\begin{equation}
\begin{cases}
\displaystyle \dot{x}_i =\alpha  x_i - \omega y_i  -\left(x_{i}^2  + y_{i}^2 \right)x_{i} + \varepsilon \sum_{j}^N A_{ij}\Big(x_{j}^3 - x_i^3 \Big),\\ \\
\displaystyle \dot{y}_i =\omega x_i + \alpha y_i - \left(x_{i}^2  + y_{i}^2 \right)y_{i}+ \varepsilon \sum_{j}^N A_{ij}\Big(x_{j}^3 - x_i^3 \Big). 
\end{cases}
\label{eq:cpn}
\end{equation}

The equations for interactions of different orders can be constructed analogously by means of the coupling functions \eqref{eq:coupl} (see also the SM of \cite{muolo2024phase}). { Before proceeding any further, let us stress the intrinsic difference between pairwise and higher-order interactions by looking at the coupling functions given by the above equations. In the pairwise setting, the clique-projected network may seem to represent a group interaction; however, in Eqs. \eqref{eq:cpn} it emerges that the units interact one-by-one, at striking difference with Eqs. \eqref{eq:3hyp}, where they interact simultaneously in a group. Stated differently, the clique-projected network represents an ensemble of interactions in pairs, while the hyperring represents a different type of interaction that involves simultaneously all the nodes of the cliques and that cannot be decomposed as the sum of pairwise interactions\footnote{{One can visualize the difference between the two settings by thinking of a social science example. Let us consider a group of four people interacting via a messaging app: a clique would represent the case in which they are writing messages to each other separately, while a hyperedge represents the interactions via a group chat.}}. }

\subsection{Hyperedge-based local order parameter}

To quantify the synchronization of an ensemble of oscillators it is common to use the Kuramoto order parameter \cite{kuramoto1975}, which gives a global measure of how much the oscillators are synchronized. However, chimera states involve the coexistence of coherent (i.e., synchronized) and incoherent (i.e., desynchronized) oscillators and the respective regions are localized. Hence, a global measure of the synchronization does not provide useful information on the chimera state. For this reason, scholars have proposed a \textit{local} Kuramoto order parameter, which measures the synchronization in a given region of the network, by quantifying the differences between neighboring oscillators, as was done, for instance, in \cite{gambuzza2016pinning}. In the case of hyperrings, a natural extension of the local order parameter would need to account for the synchronization inside each hyperedge rather than some arbitrary neighborhood. Partially inspired by a work by Shanahan \cite{shanahan2010metastable}, where the order parameter is defined with respect to the communities of a network, we hereby define a \textit{hyperedge-based local order parameter} as follows \begin{equation}
    \mathcal{R}_{i}^{\mathcal{E}}(t)=\displaystyle\Bigg|\frac{1}{\mathcal{E}_i}\sum_{j\in \mathcal{E}_i} e^{\mathrm{\imath}\vartheta_j(t)} \Bigg| ,
\end{equation}
where $\mathrm{\imath}$ is the imaginary unit, $\vartheta_j$ is the (time evolving) phase of the $j$-th oscillator and $\mathcal{E}_i$ is the hyperedge(s) node $i$ is part of. By taking as example a $3$-hyperring, Fig. \ref{fig:3hyp_vs_clique}a), we can see that, if node $i$ is a junction node, then it is part of $2$ different hyperedges and will have $6$ neighboring nodes, while non-junction nodes will have only $3$ neighbors. From that, we can observe that non-junction nodes will have the same hyperedge-based local order parameter $\mathcal{R}_{i}^{\mathcal{E}}$. The number of nodes with the same $\mathcal{R}_{i}^{\mathcal{E}}$ in each hyperedge increases with the order of the hyperring: for instance, in $3$-hyperrings they will be $2$, while in $6$-hyperrings they will be $5${, and so on.} \\ In analogy with the hyperedge-based local order parameter, for the clique-projected network we define a clique-based local order parameter as follows \begin{equation}
    \mathcal{R}_{i}^{\mathcal{C}}(t)=\displaystyle\Bigg|\frac{1}{\mathcal{C}_i}\sum_{j\in \mathcal{C}_i} e^{\mathrm{\imath}\vartheta_j(t)} \Bigg| ,
\end{equation} where $\mathcal{C}_i$ is the clique(s) node $i$ is part of.

\subsection{Pinning control}

Chimera states are often elusive patterns, emerging only for limited ranges of the parameters and specific initial conditions \cite{zakharova2020chimera}. Higher-order interactions greatly enhance the possibility of observing such a behavior \cite{ghosh_chimera2,muolo2024phase,ghosh_chimera_high-order,zhang2021unified}. Nonetheless, \textit{ad hoc} initial conditions remain a fundamental prerequisite for the chimera to emerge. Our goal is, hence, to induce chimera states in settings where they would not spontaneously appear. For this, we put to use a popular technique in control theory, called pinning control, which consists in externally acting on a subset of the nodes to drive the dynamics of the whole ensemble of nodes towards a desired state \cite{grigoriev1997pinning,sorrentino2007controllability}, and has been successfully applied in the context of chimera states on networks \cite{gambuzza2016pinning}.  In our setting, the pinning will act as a perturbation on a subset of the nodes, with the goal of developing a region of incoherence, while leaving the unperturbed nodes in their synchronous state. 
\begin{figure}[h!]
\includegraphics[scale=0.13]{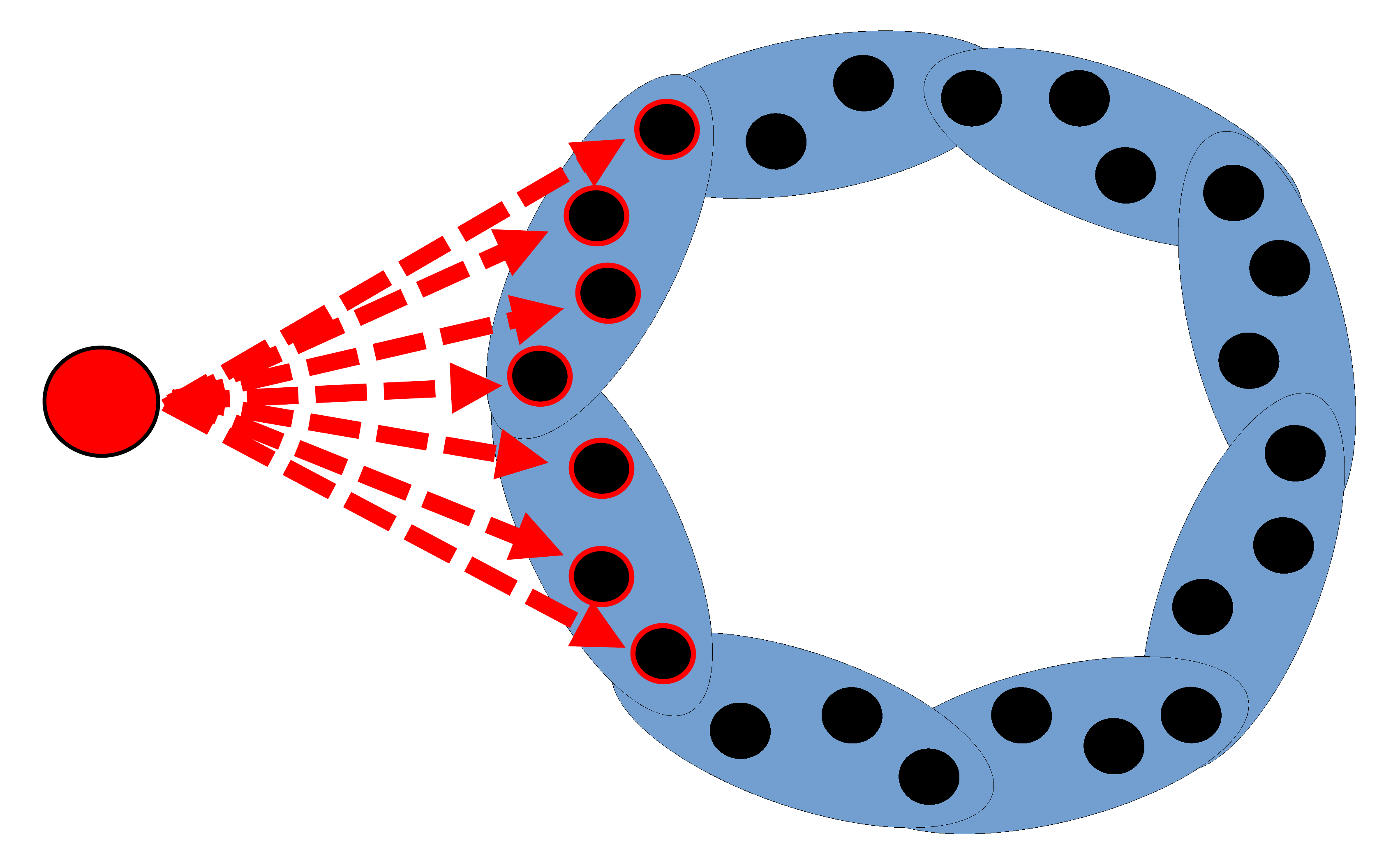}
\caption{Scheme of the pinning control for a $3$-hyperring of $24$ nodes. The system starts in the synchronized state and the control input is applied to a subset {$\mathcal{I}_p = \{1, \ldots, N_p\}$} of the nodes to induce the emergence of a chimera state.} \label{fig:pin}
\end{figure}

{In this work, we employ two distinct protocols to implement pinning control in the system, namely \emph{additive pinning} and \emph{parametric pinning}. The equations governing the controlled system under each protocol are presented in the following subsections. Here, we introduce two parameters that are common to both protocols, which serve to specify the nodes that are subject to control (i.e., pinned) and the time interval during which the control is active. Specifically, we indicate with $t_p$ the time during which the control is active, and with $N_p< N$ the number of pinned nodes. In addition, we denote the set of all nodes by $\mathcal{I} = \{1, \ldots, N\}$, and indicate the set of pinned nodes by $\mathcal{I}_p$. In most cases, we assume that the pinned nodes correspond to the first $N_p$ nodes, i.e., $\mathcal{I}_p = \{1, \ldots, N_p\}$.
The pinning control setting, under this assumption, is schematically depicted in Fig.~\ref{fig:pin}.}


\subsubsection{Control protocol I: additive pinning control}

{Let us now proceed with the first control protocol, namely \emph{additive pinning control}. As a preliminary remark, we note that this control strategy was successfully employed in~\cite{gambuzza2016pinning} to induce chimera states in systems with pairwise interactions. However, achieving such states required controlling approximately half of the nodes. In Sec.~\ref{sec:numerics}, where we present our numerical results, we highlight the significant advantage gained by applying this technique in systems with higher-order interactions.}

{To describe the additive pinning control, we first rewrite the system equations as follows} 

\begin{equation}\label{eq:additive}
    \dot{\vec{X}}_i=\vec{F}(\vec{X}_i)+\mathbb{D}\sum_{j_1,...,j_d=1}^N A_{ij_1...j_d}^{(d)}\Big(\vec{H}(\vec{X}_{j_1},...,\vec{X}_{j_d})-\vec{H}^{(d)}(\vec{X}_{i},...,\vec{X}_i)\Big) +\vec{U}_i,
\end{equation} 

\noindent {where $\vec{U}_i$ is an additive term implementing the control action. This term is given by}

\begin{equation}
{\vec{U}_i=\left \{ 
    \begin{array}{l}
    \vec{0}, \quad i \in \mathcal{I}\setminus\mathcal{I}_p  \\
    (u_{i}(t),u_{i}(t))^\top, \quad i \in \mathcal{I}_p
    \end{array}
    \right.}
\end{equation}

\noindent with

\begin{equation}
    u_{i}(t)= \lambda_{i}[\Theta(t)-\Theta(t-t_p)],
\end{equation}

\noindent where $\lambda_{i}$ are parameters drawn from a uniform distribution of a given interval and $\Theta$ is the Heaviside step function, whose value is $1$ when the argument is positive and $0$ when it is null or negative. This way, the control is active as long as $t\leq t_p$. 

{As an example, we consider the application of the protocol to the $3$-hyperring. In this case, the equations of the pinned nodes read} 

\begin{equation}
{
\begin{cases}
\displaystyle \dot{x}_i =\alpha  x_i - \omega y_i  -\left(x_{i}^2  + y_{i}^2 \right)x_{i} + \varepsilon \sum_{j_1,j_2,j_3}^N A_{ij_1j_2j_3}^{(3)}\Big(x_{j_1}x_{j_2}x_{j_3} - x_i^3\Big)+\lambda_{i}[\Theta(t)-\Theta(t-t_p)],\\ \\
\displaystyle \dot{y}_i =\omega x_i + \alpha y_i - \left(x_{i}^2  + y_{i}^2 \right)y_{i}+ \varepsilon \sum_{j_1,j_2,j_3}^N A_{ij_1j_2j_3}^{(3)}\Big(x_{j_1}x_{j_2}x_{j_3} - x_i^3 \Big)+\lambda_{i}[\Theta(t)-\Theta(t-t_p)],
\end{cases}}
\label{eq:3hypcontrolledI}
\end{equation}

\noindent{with $i \in \mathcal{I}_p$.}

{The uncontrolled nodes ($i \in \mathcal{I}\setminus\mathcal{I}_p$) are, instead, described by Eqs.~\eqref{eq:3hyp}.}

\subsubsection{Control protocol II: parametric pinning control}

{In the second type of control, which we will call \textit{parametric pinning control}, the control law acts on the pinned nodes by modifying the parameters of the dynamical system associated to each node. To formalize this protocol, we rewrite the system of equations as follows} 

\begin{equation}\label{eq:parametric}
\dot{\vec{X}}_i=\vec{F}_i(\vec{X}_i)+\mathbb{D}\sum_{j_1,...,j_d=1}^N A_{ij_1...j_d}^{(d)}\Big(\vec{H}(\vec{X}_{j_1},...,\vec{X}_{j_d})-\vec{H}^{(d)}(\vec{X}_{i},...,\vec{X}_i)\Big),
\end{equation}

\noindent {where the uncoupled (or natural) dynamics of each node, namely $\vec{F}_i$, differentiates between pinned and unpinned nodes. This term is, in fact, given by}

\begin{equation}
{\vec{F}_i=\left \{ 
    \begin{array}{l}
    \vec{F}, \quad i \in \mathcal{I}\setminus\mathcal{I}_p  \\
    (f_{p},g_{p})^\top, \quad i \in \mathcal{I}_p
    \end{array}
    \right.}
\end{equation}

\noindent {where}

\begin{equation}
{
\begin{cases}\label{eq:SL_pinned}
 f_{p}( x_{i}, y_{i}) = \alpha  x_{i}(t) - \Omega_{p,i}(t) y_{i}(t)  -\left(x_{i}^2(t)  + y_{i}^2(t) \right)x_{i}(t), \\\\
 g_{p}( x_{i}, y_{i}) = \Omega_{p,i}(t) x_{i}(t) + \alpha y_{i}(t) - \left(x_{i}^2(t)  + y_{i}^2(t) \right)y_{i}(t) . \end{cases}
}
\end{equation}

{Here, the frequencies of the controlled nodes, $\Omega_{p,i}(t)$ for $i\in \mathcal{I}_p$, are given by} \begin{equation}\label{eq:new_omega}
    {\Omega_{p,i}(t)=\left \{ 
    \begin{array}{l}
    \omega_{p,i}, \quad t\leq t_p,  \\
    \omega, \quad t>t_p     ,
    \end{array}
    \right.}
\end{equation} 
{where $\omega_{p,i}$ are new frequencies induced by the pinning; namely, we have that during the period where the control is on ($t\leq t_p$), $\Omega_{p,i}=\omega_{p,i}$, while $\Omega_{p,i}$ switches to $\omega$ as soon as the control is switched off.} 
In our numerical implementation, the new frequencies $\omega_{p,i}$ will be drawn from a uniform distribution of a given positive interval. { Note that, despite the pinning being a perturbation of the synchronous solution, in our simulations the order of magnitude of such perturbation is large, hence a linear stability analysis would not give any useful information. In fact, the latter being a local analysis, it requires small perturbations about the equilibrium state and it is not informative when the perturbations are large.} Let us also note that the control acts only on the frequency and not on the amplitude, because we have numerically verified that acting on the amplitude has no effect whatsoever. {This can be very well understood by looking at the phase reduction analysis of Sec.~4.}

{Lastly, we exemplify the parametric pinning control protocol by considering again a $3$-hyperring. The equations of the pinned nodes read} 

\begin{equation}
{
\begin{cases}
\displaystyle \dot{x}_i =\alpha  x_i - \Omega_{p,i} y_i  -\left(x_{i}^2  + y_{i}^2 \right)x_{i} + \varepsilon \sum_{j_1,j_2,j_3}^N A_{ij_1j_2j_3}^{(3)}\Big(x_{j_1}x_{j_2}x_{j_3} - x_i^3\Big),\\ \\
\displaystyle \dot{y}_i =\Omega_{p,i} x_i + \alpha y_i - \left(x_{i}^2  + y_{i}^2 \right)y_{i}+ \varepsilon \sum_{j_1,j_2,j_3}^N A_{ij_1j_2j_3}^{(3)}\Big(x_{j_1}x_{j_2}x_{j_3} - x_i^3 \Big),
\end{cases}}
\label{eq:3hypcontrolledII}
\end{equation}

\noindent{with $i \in \mathcal{I}_p$ and $\Omega_{p,i}$ as in Eq.~\eqref{eq:new_omega}.}

{Also in this case, the uncontrolled nodes ($i \in \mathcal{I}\setminus\mathcal{I}_p$) are, instead, described by Eqs.~\eqref{eq:3hyp}.}

\section{Numerical results}\label{sec:numerics}

In this section we show the numerical results of our pinning approaches. We start by {considering the case $\mathcal{I}_p=\{1,\ldots,N_p\}$, and compare the results obtained on a $3$-hyperring}, where chimera states occur with both pinning approaches, with its corresponding clique-projected network, where chimera states do not emerge. Then, {by leveraging the structure of the hyperring, we develop a different strategy to select the pinned nodes, i.e., we define a different set $\mathcal{I}_p$ based on the structure under control, and show how this} maximizes the width of the incoherence region while the number of controlled nodes remains low. For the latter, we will show the results only for additive pinning, leaving the discussion of the analogous results obtained through parametric pinning in Appendix \ref{sec:appB}. \\ { Lastly, let us point out that there is no universal consensus on when a state is chimera or not. There are some obvious features and characteristics, both quantitative and phenomenological, however, scholars may give different definition to patterns exhibiting some incoherence \cite{haugland2021changing}. Here, we will call \textit{chimeras} the states where the incoherence regions exhibit the chaotic behavior shown by Kuramoto and Battogtokh \cite{kuramoto2002coexistence} and, for our specific case of phase chimeras, by Zajdela and Abrams \cite{zajdela_abrams}. We will call states exhibiting a weak incoherence \textit{weak chimera-like states}, inspired by the definition given by Aswhin and Burylko for chimera states with respect to frequencies \cite{ashwin2015weak}.}

\subsection{Comparison between higher-order and pairwise interactions}

\begin{figure}[h!]
\includegraphics[scale=0.29]{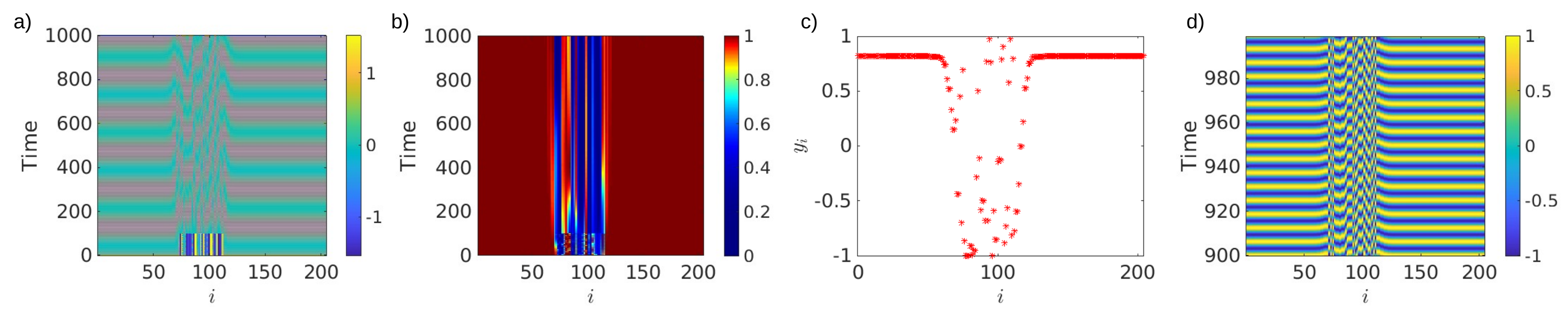}
\caption{Additive pinning induces phase chimera states on a $3$-hyperring (i.e., $4$-body interactions) of $204$ nodes. Panel a) depicts the whole time series of variables $y_i(t)$ with $i=1,...,N$, panel b) the hyperedge-based local order parameter, panel c) a snapshot of variables $y_i(t)$ with $i=1,...,N$ for $t_{final}=1000$ time units and panel d) shows a zoom of the time series of variables $y_i(t)$ with $i=1,...,N$. { The results for the variables $x_i(t)$ are analogous and, hence, not shown.} The model parameters are $\alpha=1$ and $\omega=1$ and the coupling strength is $\varepsilon=0.01$. Pinning control is applied to $N_p=40$ consecutive nodes for $t_p=100$ time units. The parameters $\lambda_{i_p}$ are drawn from a uniform distribution in the interval $[-2,2]$.} \label{fig:additive}
\end{figure}

\begin{figure}[h!]
\includegraphics[scale=0.29]{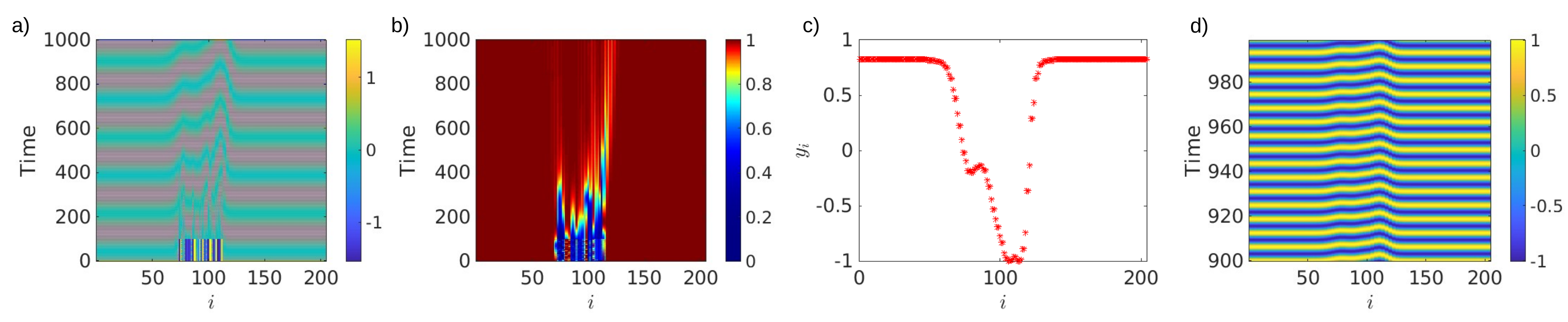}
\caption{Additive pinning on a clique-projected network of $204$ nodes. Chimera states do not emerge in this setting, at contrast with the previous Fig. \ref{fig:additive}. Panel a) depicts the whole time series of variables $y_i(t)$ with $i=1,...,N$, panel b) the clique-based local order parameter, panel c) a snapshot of variables $y_i(t)$ with $i=1,...,N$ for $t_{final}=1000$ time units and panel d) shows a zoom of the time series of variables $y_i(t)$ with $i=1,...,N$. { The results for the variables $x_i(t)$ are analogous and, hence, not shown.} The model parameters are $\alpha=1$ and $\omega=1$ and the coupling strength is $\varepsilon=0.01$. Pinning control is applied to $N_p=40$ consecutive nodes for $t_p=100$ time units. The parameters $\lambda_{i_p}$ are the same of the previous figure.} \label{fig:additive_pairwise}
\end{figure}

\begin{figure}[h!]
\includegraphics[scale=0.29]{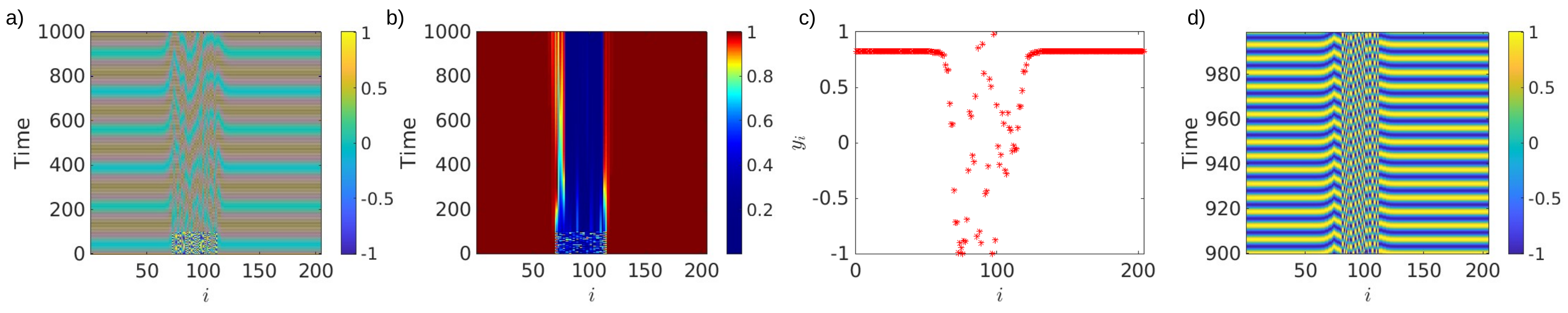}
\caption{Parametric pinning induces phase chimera states on a $3$-hyperring (i.e., $4$-body interactions) of $204$ nodes. Panel a) depicts the whole time series of variables $y_i(t)$ with $i=1,...,N$, panel b) the hyperedge-based local order parameter, panel c) a snapshot of variables $y_i(t)$ with $i=1,...,N$ for $t_{final}=1000$ time units and panel d) shows a zoom of the time series of variables $y_i(t)$ with $i=1,...,N$. { The results for the variables $x_i(t)$ are analogous and, hence, not shown.} The model parameters are $\alpha=1$ and $\omega=1$ and the coupling strength is $\varepsilon=0.01$. Pinning control is applied to $N_p=40$ consecutive nodes for $t_p=100$ time units. The parameters $\omega_{i_p}$ are drawn from a uniform distribution in the interval $[0.5,2.5]$.} \label{fig:parametric}
\end{figure}

\begin{figure}[h!]
\includegraphics[scale=0.29]{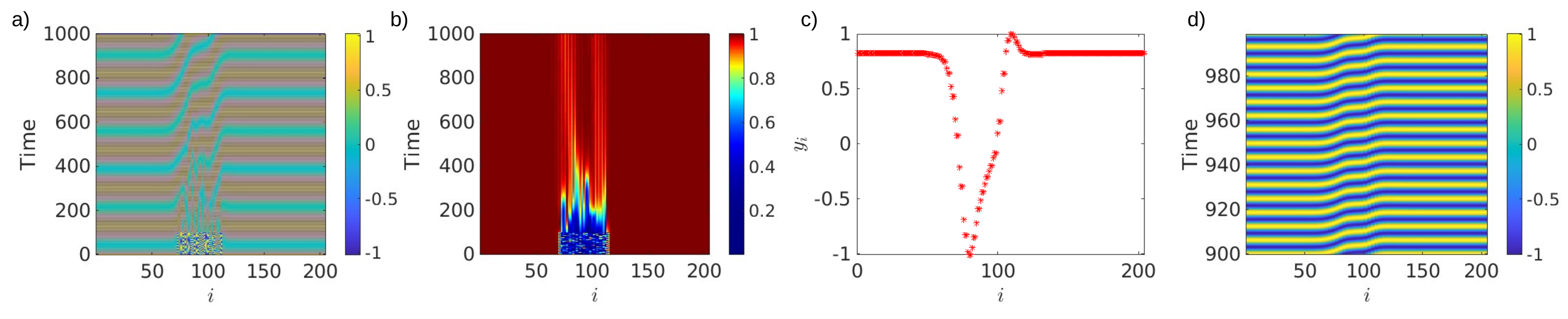}
\caption{Parametric pinning on a clique-projected network of $204$ nodes. Chimera states do not emerge in this setting, at contrast with the previous Fig. \ref{fig:parametric}. Panel a) depicts the whole time series of variables $y_i(t)$ with $i=1,...,N$, panel b) the clique-based local order parameter, panel c) a snapshot of variables $y_i(t)$ with $i=1,...,N$ for $t_{final}=1000$ time units and panel d) shows a zoom of the time series of variables $y_i(t)$ with $i=1,...,N$. { The results for the variables $x_i(t)$ are analogous and, hence, not shown.} The model parameters are $\alpha=1$ and $\omega=1$ and the coupling strength is $\varepsilon=0.01$. Pinning control is applied to $N_p=40$ consecutive nodes for $t_p=100$ time units. The parameters $\omega_{i_p}$ are the same of the previous figure.} \label{fig:parametric_pairwise}
\end{figure}

Let us proceed in testing our pinning approaches to control the emergence of chimera states on a $3$-hyperring and compare it with the clique-project network case. In Figs. \ref{fig:additive} and \ref{fig:additive_pairwise}, we show the results for the additive pinning on a hyperring and a clique-projected network, respectively, while, in Figs. \ref{fig:parametric} and \ref{fig:parametric_pairwise}, we show the results for the parametric pinning on a hyperring and a clique-projected network, respectively. {Here we always assume $\mathcal{I}_p=\{1,\ldots,N_p\}$. Note that the time series is depicted for the $y$ variables of the SL model, but the results are qualitatively the same if we display the $x$ variable; indeed, the local order parameter is computed using both variables.} For both pinning approaches, we see that the control induces a chimera state when the topology is higher-order (Figs. \ref{fig:additive} and \ref{fig:parametric}). This can be visualized through the hyperedge-{based} local order parameter $\mathcal{R}_{i}^{\mathcal{E}}$, which shows that the nodes {that} have been controlled are not oscillating coherently with respect to their neighbors sharing the same hyperedge(s) and that such incoherence persists. On the other hand, the pairwise case does not yield chimeras (Figs. \ref{fig:additive_pairwise} and \ref{fig:parametric_pairwise}). In fact, the initial decoherence induced by the control is quickly reabsorbed by the system and, although a clear trace of the pinning remains, the difference between the phases of adjacent oscillators is small and the variation smooth. This can be visualized through the clique-based order parameter $\mathcal{R}_{i}^{\mathcal{C}}$. {Let us point out that clique-projected networks obtained from hyperrings of different orders do not significantly affect the behavior of the system. Here, we have shown the results on a clique-projected network as depicted in Fig. \ref{fig:3hyp_vs_clique}, but qualitatively similar time series are obtained on different clique-projected networks, which is consistent with the results obtained in a previous work \cite{muolo2024phase}. Moreover, when the nonlinear coupling have the same functional form, the results on different clique-projected networks are the same.} Note that in \cite{muolo2024phase} {the dynamics on clique-projected networks} was distinguished from a chimera one through the total phase variation. Our approach based on a local order parameter is complementary. 

{ For the chosen parameters, the higher-order setting always yields chimera states when the consecutive pinned nodes are about $10$ and above. For lower numbers of pinned nodes, we will always have at least a weak chimera-like state, while the emergence of chimera states depends on the random perturbation of the pinning. Note that, when the number of pinned nodes is very small (e.g., $1\sim 3$), the resulting pattern will always be a weak chimera-like state. Moreover, such weak chimera-like state persists for long integration times. In Figs. \ref{fig:additive} and \ref{fig:parametric}, we show the temporal evolution until $1000$ time units, while the chimera state persists until about $3000 \sim 4000$ time units (depending on the perturbation of the pinning). After such integration time, the chimera state turns into a weak chimera-like state, where variation between adjacent phases is smoother, but $\mathcal{R}_{i}^{\mathcal{E}}$ remains low. We have found that the weak chimera state persists with  almost constant $\mathcal{R}_{i}^{\mathcal{E}}$ until $20000$ time units (the maximum integration time tested). This is particularly striking when compared to the behavior observed on clique-projected networks, where the incoherence vanishes after a short time. To better appreciate the difference between the higher-order and pairwise cases, we provide $4$ Supplementary Videos (SVs), where the dynamics until $5000$ time units can be visualized: SV1 is obtained from a setting analogous to Fig. \ref{fig:additive} (i.e., additive pinning on a $3$-hyperring), SV2 is obtained from a setting analogous to Fig. \ref{fig:additive_pairwise} (i.e., additive pinning on a clique-projected network), SV3 is obtained from a setting analogous to Fig. \ref{fig:parametric} (i.e., parametric pinning on a $3$-hyperring), and SV4 is obtained from a setting analogous to Fig. \ref{fig:parametric_pairwise} (i.e., parametric pinning on a clique-projected network).}

Additionally, let us stress that all results hereby shown, regardless of the order of the hyperring, coupling and pinning approach, are due to the higher-order topology and no chimera states are found when performing the simulations with the same setting but on the corresponding clique-projected network, exactly as in the figures shown in this section. Note, though, that there is one particular exception discussed in Appendix \ref{sec:appA}, where the observed pattern is not due to the higher-order topology but due to the coupling configuration, and, in fact, it is found also in the corresponding pairwise system. Such results provide further evidence that higher-order interactions promote the presence of chimera states and are consistent with the existing literature \cite{ghosh_chimera2,muolo2024phase,ghosh_chimera_high-order,zhang2021unified}.

Let us conclude by pointing out that, in previous works, chimera states were obtained for specific values of the initial conditions, while random or uniform initial conditions did not yield the same result. In our numerical study, the initial conditions do not matter as long as the system starts on the synchronous solution, i.e., on the limit cycle of the Stuart-Landau oscillator \eqref{eq:SL}. Hence, we will choose the initial conditions to be uniform and without noise, i.e., $(x_j^0=1,y_j^0=0)$ for every oscillator.

\subsection{Scaling of the pinned subset with respect to the hyperring size}

\begin{figure}[h!]
\includegraphics[scale=0.12]{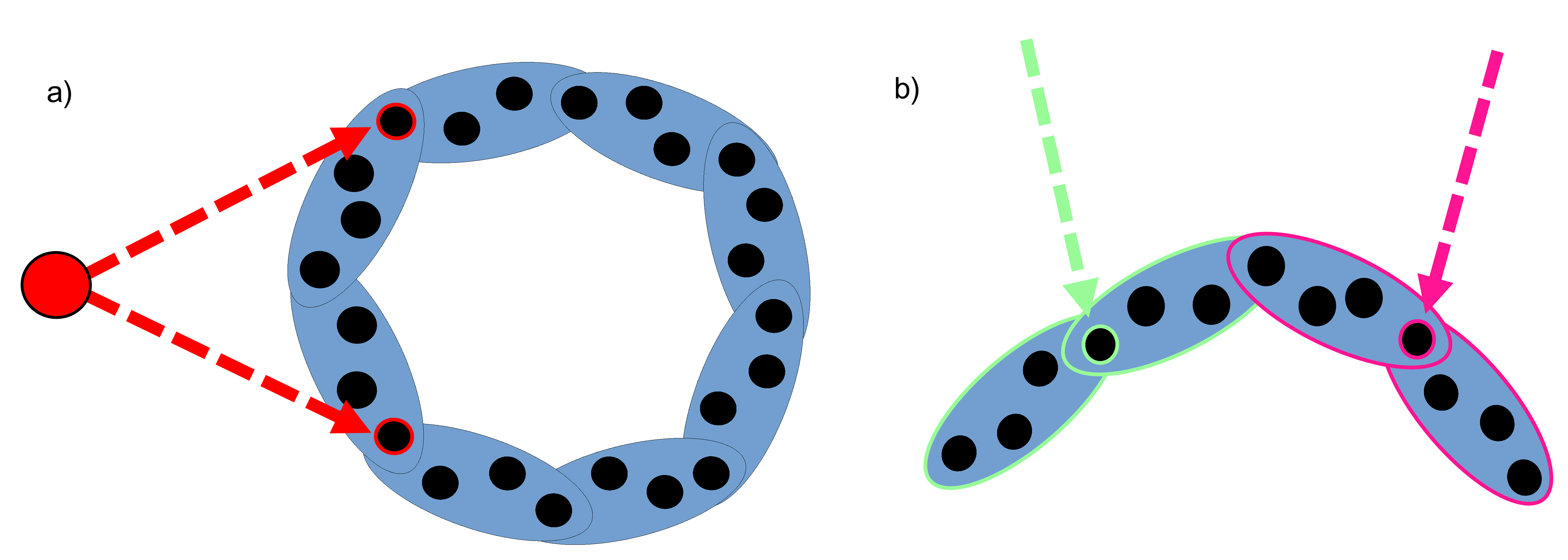}
\caption{Pinning scheme allowing us to exploit the structure of the hyperring. In panel a), we show a setting in which we can, in principle, control half of the nodes while pinning only $2$ nodes. In panel b), it is schematically shown how a control of the junction node affects two hyperedges. Such protocol, shown for a $3$-hyperring, is straightforwardly extended to any $d$-hyperring.} \label{fig:pinning_scheme}
\end{figure}

 {The previous setting of consecutively pinned nodes exhibit some features that are independent of the size but are caused by the sole structure of the topology. In fact, independently of the total size of the hyperring and of the number of pinned nodes, the nodes which are affected by the perturbation outside of the pinning region are also of the order of $2$ hyperedges. Hence, the chimera state will develop on the pinned region with a boundary of about two hyperedges. We can exploit this property of the topology to obtain chimera states with a very small percentage of pinned nodes.}
 Hence, we set up a pinning protocol in which we try to maximize the number of nodes affected by one single controller. Due to the hyperring structure, a way can be to pin every other junction node, so that each controlled node can, in principle, affect two hyperedges, as schematically shown in Fig. \ref{fig:pinning_scheme}. To observe how the number of pinned nodes scales with the size of the hyperring, we keep constant the number of hyperedges, so that the total number of nodes increases, but the backbone of the structure remains unchanged. Given that each pinned node is part of $2$ hyperedges, in principle, we are able to control all the nodes in these hyperedges. E.g., in a $3$-hyperring, with each pinned nodes we would control $7$ nodes, in a $4$-hyperring $9$ nodes, in a $5$-hyperring $11$ nodes and in a $6$-hyperring $13$ nodes. For brevity, we present here the results for the additive pinning. The results for the parametric pinning are similar and can be found in Appendix \ref{sec:appB}. 

\begin{figure}
\includegraphics[scale=0.3]{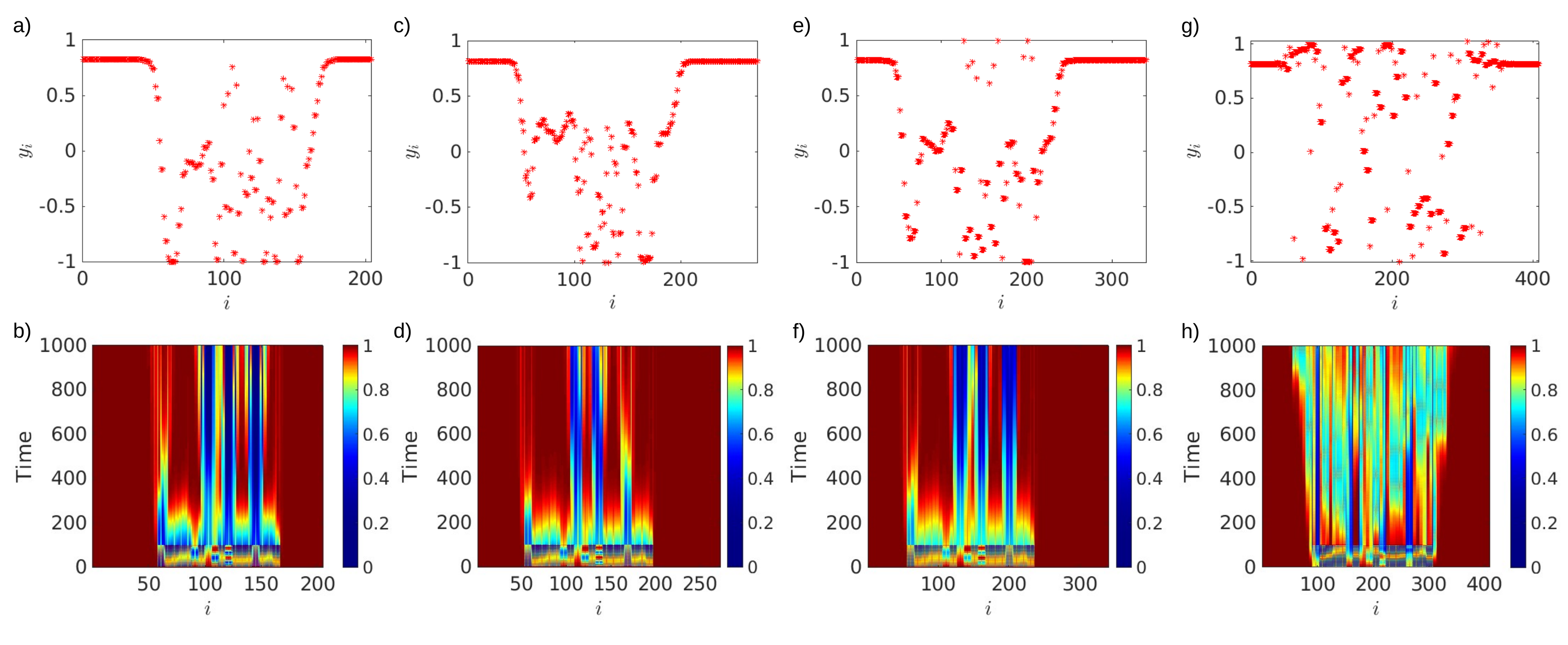}
\caption{Scaling of the pinned subset with respect to the higher-order structure size with the additive pinning approach. On the upper panels we depict the snapshots of variables $y_i(t)$ with $i=1,...,N$ for $t_{final}=1000$ time units, while the lower panels show the hyperedge-based local order parameters. { The results for the variables $x_i(t)$ are analogous and, hence, not shown.} Panels a) and b) report the simulations performed on a $3$-hyperring (i.e., $4$-body interactions) of $204$ nodes, panels c) and d) on a $4$-hyperring (i.e., $5$-body interactions) of $272$ nodes, panels e) and f) on a $5$-hyperring (i.e., $6$-body interactions) of $340$ nodes and panels g) and h) on a $6$-hyperring (i.e., $7$-body interactions) of $408$ nodes. The number of nodes is chosen so that each hyperring is made of $68$ hyperedges. Pinning control is applied to $N_p=18$ nodes as in Fig. \ref{fig:pinning_scheme}, i.e., every two junction nodes, to all the structures and for $t_p=100$ time units. This means that the pinned nodes are $\simeq 8.8\%$ of the total nodes in the $3$-hyperring, $\simeq 6.6\%$ of the total nodes in the $4$-hyperring, $\simeq 5.3\%$ of the total nodes in the $5$-hyperring and $\simeq 4.4\%$ of the total nodes in the $6$-hyperring. The model parameters are $\alpha=1$ and $\omega=1$, the coupling strength is $\varepsilon=0.01$, except for the $6$-hyperring where $\varepsilon=0.1$, and the parameters $\lambda_{i_p}$ are drawn from a uniform distribution in the interval $[0,2]$ and are the same for all the simulations.} \label{fig:scaling_add}
\end{figure}

In Fig. \ref{fig:scaling_add}, we show the results obtained with such control scheme on $d$-hyperrings with $d=3,4,5,6$, i.e., $4$-, $5$-, $6$- and $7$-body interactions, where we have fixed the number of hyperedges. Indeed, we can observe that we obtain a chimera state by inducing a large region of incoherence (more than half of the nodes) with a control that involves only a small fraction of the nodes. Moreover, the pinned nodes {$\mathcal{I}_p = \{1, \ldots, N_p\}$} are kept constant for every structure, meaning that $N_p$ does not scale with the number of nodes, but rather with the number of hyperedges, which allows to control large structures with only a handful of nodes. The pinned nodes are $\simeq 8.8\%$ of the total nodes for the $3$-hyperring (panels a) and b)), $\simeq 6.6\%$ of the total nodes for the $4$-hyperring (panels c) and d)), $\simeq 5.3\%$ of the total nodes in the $5$-hyperring (panels e) and f)) and $\simeq 4.4\%$ of the total nodes in the $6$-hyperring (panels g) and h)). In the latter case, our pinning scheme does not work as well as for the other structures and the coupling strength {needs} to be significantly increased in order to observe a chimera state. However, we can observe from Fig. \ref{fig:scaling_add}h) that the chimera is not anymore stable and the front of incoherence enlarges. Let us stress that stable chimera states can be easily obtained also in $6$-hyperrings by reducing the distance between the controlled nodes, as in the previous section, instead of the pinning scheme of Fig. \ref{fig:pinning_scheme}.

Let us note that{, for the pinning scheme of Fig. \ref{fig:pinning_scheme},} the parameters $\lambda_{i_p}$ need to be all positive or all negative in order for this pinning scheme to yield persistent chimera states. When such random inputs are drawn in a symmetric interval with respect to the $0$, e.g., $[-2,2]$ as in Fig. \ref{fig:additive}, a chimera states forms but {turns into a weak chimera-like state} at about $400$ time units. On the other hand, when the control inputs have the same sign, e.g., $[0,2]$ as in Fig. \ref{fig:scaling_add}, the chimera is persistent until about $4000$ time units, i.e., $10$ times longer{, before turning into a weak chimera-like state}.

In Appendix \ref{sec:appB}, we show the results for the parametric pinning, that are analogous except for the case {of} $6$-hyperring, where {only weak chimera-like states emerge when pinning every other junction node,} even with stronger couplings. The fact that our results are robust with respect to different control approaches is a good indication of the pervasiveness of the phenomenon. 

Let us conclude the Results Section by pointing out that the chimera behavior can be further enhanced by increasing the number of pinned nodes and/or reducing the distance between them. Moreover, by increasing the magnitude of the parameters $\lambda_{i_p}$, we can also obtain chimera states through controlling even less nodes than in the simulations hereby shown. However, we have presented a setting in which the parameters $\lambda_{i_p}$ have a magnitude comparable with the involved parameters, in order to make it suitable for applications.

\section{Heuristic interpretation through phase reduction theory}\label{sec:heuristic}

In this section we will give a heuristic interpretation of the results based on the phase reduction approach \cite{nakao2016phase,kuramoto2019concept}, which consists of reducing a given oscillatory system to a phase, i.e., Kuramoto-type, model \cite{kuramoto1975}. In a nutshell, starting from a system of $N$ highly dimensional units in a limit cycle regime, e.g., \begin{equation}
    \dot{\vec{X}}_i=\vec{F}_i(\vec{X}_i)+\varepsilon\sum_{j=1}^N A_{ij}\vec{G}_{ij}(\vec{X}_j,\vec{X}_i),
\end{equation} under given assumptions (see \cite{nakao2016phase} for details), we can reduce to a system of $N$ interacting phase oscillators of the form \begin{equation}
        \dot{\vartheta}_i=\omega_i+\varepsilon\sum_{j=1}^N A_{ij}\vec{Z}(\vartheta_i)\cdot\vec{G}_{ij}(\vartheta_j,\vartheta_i),
\end{equation} where $\omega_i$ is the frequency of the $i$-th oscillator and $\vec{Z}$ is the phase sensitivity function. The key in the reduction process is to find an expression for $\vec{Z}$, which {is analytical only in} some specific cases, among which the Stuart-Landau model \cite{nakao2016phase} and weakly nonlinear oscillators \cite{leon2023analytical}, while it needs to be obtained numerically for general oscillators. \\

The phase reduction theory has been applied also to systems with higher-order interactions \cite{leon2025theory,ashwin2016hopf,leon2019phase,leon2024higher}, obtaining higher-order versions of the Kuramoto model, which exhibit much richer behaviors than the pairwise one. Indeed, the first evidence of higher-order-induced exotic behaviors, which triggered the interest of the community towards this new framework, has come from higher-order phase models (although not derived through phase reduction) \cite{tanaka2011multistable,bick2016chaos,skardal2019abrupt,skardal2019higher,millan2020explosive}. In what follows, we will apply the phase reduction to our model on a $3$-hyperring and on the corresponding clique-projected network, Eqs. \eqref{eq:3hyp} and \eqref{eq:cpn} respectively, to give an intuition of why it is easier to induce chimera behavior via pinning when the topology is higher-order. \\
For the Stuart-Landau model, the phase sensitivity function is $\vec{Z}(\vartheta)=(-\sin(\vartheta),\cos(\vartheta))^\top$ \cite{nakao2016phase}. Let us consider system \eqref{eq:3hyp} in polar coordinates, i.e., $\vec{X}_i=(\cos(\vartheta_i),\sin(\vartheta_i))$, and proceed with the reduction by computing the following \begin{eqnarray}
    \vec{Z}(\vartheta_i)\cdot \dot{\vec{X}}_i &=& \sin^2(\vartheta_i)\dot{\vartheta_i}+\cos^2(\vartheta_i)\dot{\vartheta}_i= \nonumber \\ &&-\alpha\cos(\vartheta_i)\sin(\vartheta_i)+\omega\sin^2(\vartheta_i)+\cos(\vartheta_i)\sin(\vartheta_i)+\omega\cos^2(\vartheta_i)+\alpha\cos(\vartheta_i)\sin(\vartheta_i) \nonumber \\ && -\cos(\vartheta_i)\sin(\vartheta_i)+ \varepsilon\sum_{j_1,j_2,j_3}^N A_{ij_1j_2j_3}^{(3)}\Big( \cos^3(\vartheta_i)\sin(\vartheta_i)- \nonumber \\ &&\cos(\vartheta_{j_1})\cos(\vartheta_{j_2})\cos(\vartheta_{j_3})\sin(\vartheta_i)+\cos(\vartheta_{j_1})\cos(\vartheta_{j_2})\cos(\vartheta_{j_3})\cos(\vartheta_i)-\cos^4(\vartheta_i) \Big), \nonumber
\end{eqnarray} which gives \begin{equation}\label{eq:phase_higher}
    \dot{\vartheta}_i=\omega+\varepsilon\sum_{j_1,j_2,j_3}^N A_{ij_1j_2j_3}^{(3)} \Phi( \vartheta_i, \vartheta_{j_1}, \vartheta_{j_2}, \vartheta_{j_3} ),
 \end{equation} where \begin{eqnarray}\label{eq:4_naveraged}
 \Phi( \vartheta_i, \vartheta_{j_1}, \vartheta_{j_2}, \vartheta_{j_3} )&=&  \cos^3(\vartheta_i)\sin(\vartheta_i)-\cos(\vartheta_{j_1})\cos(\vartheta_{j_2})\cos(\vartheta_{j_3})\sin(\vartheta_i)+\nonumber \\ &&\cos(\vartheta_{j_1})\cos(\vartheta_{j_2})\cos(\vartheta_{j_3})\cos(\vartheta_i)-\cos^4(\vartheta_i).
\end{eqnarray}
Observe that $\omega$ is the same for all the oscillators because we started from identical Stuart-Landau systems. If we apply the same procedure to the system on the clique-projected network, i.e., Eq. \eqref{eq:cpn}, we obtain 

\begin{equation}\label{eq:phase_pairwise}
        \dot{\vartheta}_i=\omega+\varepsilon\sum_{j}^N A_{ij}\Psi( \vartheta_i, \vartheta_{j} ),
\end{equation}
where \begin{equation}\label{eq:naveraged}
\Psi( \vartheta_i, \vartheta_{j} ) =\big(\cos(\vartheta_i)-\sin(\vartheta_i)\big)\big(\cos^3(\vartheta_j)-\cos^3(\vartheta_i)\big). \end{equation} 

Through the averaging method \cite{nakao2016phase}, the $4$-body coupling \eqref{eq:4_naveraged} can be approximated as \begin{equation}\label{eq:4_averaged}
\Phi( \vartheta_i, \vartheta_{j_1}, \vartheta_{j_2}, \vartheta_{j_3} ) \simeq \frac{3}{8} \Big( \sin(\vartheta_{j_1}+\vartheta_{j_2}+\vartheta_{j_3}-3\vartheta_i) + \cos(\vartheta_{j_1}+\vartheta_{j_2}+\vartheta_{j_3}-3\vartheta_i)  -1\Big), \end{equation} while the pairwise coupling as \eqref{eq:naveraged} 
\begin{equation}\label{eq:averaged}
 \Psi( \vartheta_i, \vartheta_{j} )  \simeq \frac{3}{8}\Big( \sin(\vartheta_j-\vartheta_i)+\cos(\vartheta_j-\vartheta_i)-1 \Big)= \sqrt{2}\frac{3}{8}\sin\Big(\vartheta_j-\vartheta_i+\frac{\pi}{4}\Big)-\frac{3}{8}.
\end{equation}
 
 The coupling given by Eq. \eqref{eq:averaged} steers the system towards synchronization, because $\Psi( \vartheta_i, \vartheta_{j} )$ has form of the well-known Kuramoto-Sakaguchi coupling, i.e., $\sin(\vartheta_j-\vartheta_i+\alpha)$, which is known to be attractive for $|\alpha|<\frac{\pi}{2}$ \cite{Kuramoto}. On the other hand, Eq. \eqref{eq:4_averaged} allows for a much richer dynamics, given that there are many more combinations of the phases and the coefficients such that the coupling term vanishes, as it is the case of the higher-order Kuramoto model \cite{leon2024higher}. This fact gives an intuition not only of the much richer dynamics observed when higher-order interactions are present \cite{tanaka2011multistable,skardal2019higher,millan2020explosive,leon2024higher,skardal2019abrupt}, but also of why higher-order systems favor the presence of chimera states and it is much easier to induce such behavior via pinning, compared to the pairwise case. Moreover, given the form of the higher-order coupling terms, once such state is achieved, it is more difficult for the higher-order system to steer towards synchronization, which provides a qualitative explanation of why the chimera states are also persistent. Let us remark that the first intuition of this behavior was given in \cite{zhang2024deeper}, where it was shown, for the higher-order Kuramoto model, that, when the system leaves the attraction basin of the synchronous state, it is more difficult to synchronize again because higher-order interactions cause a shrinking of such attraction basin, which becomes smaller. {A similar conclusion has been provided also in \cite{von2024higher}.} In our case, the control pushes the system away from the synchronous solution creating a chimera state and the higher-order interactions favor the persistence of this state. 

 \begin{figure}
\includegraphics[scale=0.29]{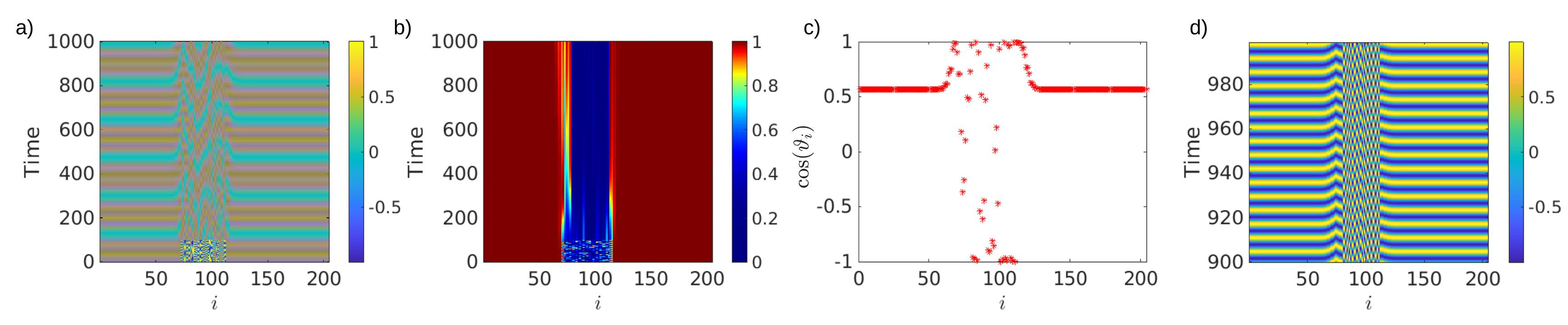}
\caption{Pinning induces phase chimera states for the reduced phase model \eqref{eq:phase_higher} on a $3$-hyperring (i.e., $4$-body interactions) of $204$ nodes. Panel a) depicts the whole time series of {the cosine of} the phases {$\cos(\vartheta_i(t))$} with $i=1,...,N$, panel b) the hyperedge-based local order parameter, panel c) a snapshot of the {cosine of the} phases {$\cos(\vartheta_i(t))$} with $i=1,...,N$ for $t_{final}=1000$ time units and panel d) shows a zoom of the time series of the {cosine of the} phases {$\cos(\vartheta_i(t))$} with $i=1,...,N$. The frequency is $\omega=1$ and the coupling strength is $\varepsilon=0.01$. Pinning control is applied to $N_p=40$ consecutive nodes for $t_p=100$ time units. The parameters $\lambda_{i_p}$ are the same of Figs. \ref{fig:additive} and \ref{fig:additive_pairwise}.} \label{fig:additive_phase}
\end{figure}

\begin{figure}
\includegraphics[scale=0.29]{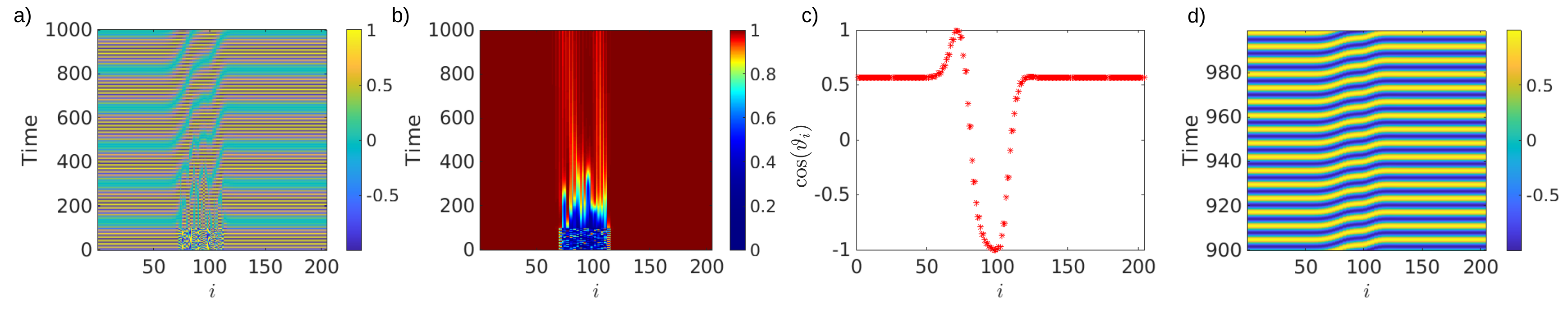}
\caption{Pinning for the reduced phase model \eqref{eq:phase_pairwise} on a clique-projected network of $204$ nodes. Chimera states do not emerge in this setting, at contrast with the previous Fig. \ref{fig:additive_phase}. Panel a) depicts the whole time series of {the cosine of} the phases {$\cos(\vartheta_i(t))$} with $i=1,...,N$, panel b) the clique-based local order parameter, panel c) a snapshot of {the cosine of} the phases {$\cos(\vartheta_i(t))$} with $i=1,...,N$ for $t_{final}=1000$ time units and panel d) shows a zoom of the time series of {the cosine of} the phases {$\cos(\vartheta_i(t))$} with $i=1,...,N$. The frequency is $\omega=1$ and the coupling strength is $\varepsilon=0.01$. Pinning control is applied to $N_p=40$ consecutive nodes for $t_p=100$ time units. The parameters $\lambda_{i_p}$ are the same of Figs. \ref{fig:additive} and \ref{fig:additive_pairwise}.} \label{fig:additive_pairwise_phase}
\end{figure}

As a further corroboration of the results shown in this work and of the correctness of our phase reduction approach, let us conclude this section by applying our pinning protocols to the reduced phase models and show that the outcome confirms our claims. { Before proceeding with the implementation of the pinning protocol, let us observe that for a phase model, such as that of Eq. \eqref{eq:phase_higher}, the additive and parametric protocols are equivalent. In fact, the uncoupled oscillators are described by the sole frequency $\omega$, hence, an additional input (i.e., additive pinning) or a modification of the latter (parametric pinning) are the same operation. It is important to note that this does not mean that the two pinning protocols are always equivalent, but, when the coupling is small enough for the phase reduction to be valid, the approximated phase model obtained after averaging reacts in the same way for both protocols. In Figs. \ref{fig:additive_phase} and \ref{fig:additive_pairwise_phase}, we show the result of the pinning procedure on a $3$-hyperring and its corresponding clique-projected network, respectively. The results are analogous to those of the non-reduced model: while the control induces a chimera state in the higher-order phase reduced model, no chimeras emerge in the pairwise setting. This was expected, given that we are dealing with phase chimeras and, hence, the phase model should fully reproduce such patterns. Again, note that the phase reduction is an approximation, and the phase reduced model does not behave exactly as the full model, but there may be small differences\footnote{{These differences can be particularly relevant close to bifurcation points \cite{leon2019phase,bick2024higher}, but this is not the case in the setting we consider here.}}.}

\section{Discussion}

In this work we have shown how pinning control can be applied to higher-order {systems} to trigger the emergence of chimera states and how higher-order interactions are a key feature for the chimera state to develop and persist. It was already known from previous works that higher-order interactions enhanced chimera states, however, the set of parameters, initial conditions and couplings allowing for such behavior remained limited. Thanks to our pinning schemes, that we called additive pinning and parametric pinning, we were able to overcome such limitations and observe chimera patterns for a wide range of settings. Moreover, and this is the most remarkable result, the higher-order framework makes possible to control the presence of chimeras by only acting on a small fraction of the nodes, at striking contrast with the network case, where {about} half of the nodes {need} to be controlled to achieve this objective. Lastly, our heuristic interpretation of the results goes beyond this work and provides a possible explanation of other previous results regarding synchronization patterns observed in higher-order systems, in particular the claim made by Zhang et al. that higher-order interactions shrink the attraction basin of the synchronized state \cite{zhang2024deeper}.

Our results clearly show that it is easier and more efficient to trigger the emergence of chimera states when higher-order interactions are present. A further study could be to determine how much energy is needed to control the chimera state in comparison with the pairwise setting, by relying on energy aware controllability measures \cite{lindmark2018minimum,baggio2022energy}. { Indeed, in \cite{zhang2024deeper}, as well as in other works such as \cite{von2024higher}, it is proven that the synchronous state is more robust to perturbations in the presence of higher-order interactions. This means that, in principle, it may be more costly, in terms of energy, to desynchronize coupled oscillators on hypergraphs. A further direction of study could be an estimation from the energy needed to obtain chimera states in our setting. They are easier to obtain, with respect to the case of networks, and with a longer life time, but how costly are they? This question may be very challenging, because the main energy estimation methods developed in control theory consist in computing the energy needed to steer the system onto a desired trajectory from a given initial state \cite{baggio2022energy}; in the case of chimera states, we do not have an expression of the desired trajectory, not even numerically, making the estimation problem difficult.} Another efficient control strategy could be to apply an intermittent pinning, analogously to the occasional coupling setting developed in the framework of amplitude death \cite{ghosh2022occasional}, where techniques from piecewise-smooth systems could be used \cite{coraggio2021convergence}. Another interesting direction would be to apply pinning control to directed higher-order structures, such as directed \cite{gallo1993directed} and $m$-directed hypergraphs \cite{gallo2022synchronization}, which have been proven to greatly affect nonlinear dynamics in the context of synchronization \cite{della2023emergence} and Turing pattern formation \cite{dorchain2024impact}, respectively. Pinning approaches in systems with directed higher-order interactions have been developed in some pioneering works \cite{della2023emergence,shi2023synchronization,li2024synchronization,rizzello2024pinning}, but not yet in the context of chimera states. { Lastly, the property of higher-order structures to conserve the incoherence could be exploited by engineering the perturbation to attain a desired incoherent state. A possible application could be in the framework of electric circuits, where higher-order interactions have been recently implemented \cite{vera2024electronic,minati2024chaotic}.} \\

In conclusion, this work is one of the {firsts} in which methods from control theory are applied to systems with higher-order interactions and it shows the numerous possibilities offered by this novel framework. We believe that there is plenty of exciting research to be done in this direction and that the ground we built with this work {sets} the basis for further studies shedding more light on the interplay between dynamics with higher-order interactions and its control.

\begin{acknowledgements}
The work of R.M. is supported by a JSPS postdoctoral fellowship, grant 24KF0211. H.N. acknowledges JSPS KAKENHI 25H01468, 25K03081, and 22H00516 for financial support. The contribution of L.V.G. and M.F. to this work is framed into the activities of the project "CoCoS: Control of Complex Systems", funded by the University of Catania, under the PIA.CE.RI. 2024-26 initiative. We are grateful to Timoteo Carletti for useful discussions, including pointing us to Ref. \cite{shanahan2010metastable}. R.M. is particularly grateful to Martin Moriamé, Giacomo Baggio and Francesco Lo Iudice for the discussions on pinning control and to Iván León and Yuzuru Kato for the discussions on the phase reduction.
\end{acknowledgements}

%
\section*{Conflict of interest}
 The authors declare that they have no conflict of interest.

\section*{Author contributions} Conceptualization: R.M., L.V.G., M.F.; Methodology: R.M., L.V.G., H.N., M.F.; Software: R.M.; Formal Analysis: R.M., H.N.; Validation: R.M.; Writing-Original Draft: R.M.; Writing-Review and Editing: R.M., L.V.G., H.N., M.F.; Funding Acquisition: H.N., M.F.


%


\appendix

\section{Numerical results for different coupling schemes}
\setcounter{equation}{0}
\renewcommand{\theequation}{A\arabic{equation}}
\setcounter{figure}{0}
\renewcommand{\thefigure}{A\arabic{figure}}\label{sec:appA}

In the Main Text, we have considered systems of the form of Eq. \eqref{eq:SL_HO}, where the coupling matrix is $D=\begin{bmatrix} 1& 0 \\ 1 & 0 \end{bmatrix}$. Let us observe that such setting is the one in which it is easier to observe chimera states induced by the initial conditions. However, through our pinning approach, it is possible to observe chimera states also for different configurations of the coupling.  In what follows, we give a brief survey of which configurations yield chimera states, obtained by performing parametric pinning control on a $3$-hyperring of $204$ nodes, where the parameters are $\alpha=1$ and $\omega=1$, the coupling strength is $\varepsilon=0.01$, and by pinning $N_p=18$ nodes spaced every $3$ for $t_p=100$ time units and with the parameters $\omega_{i_p}$ drawn from a uniform distribution of the interval $[0.5,2.5]$. We have performed $10$ simulations for each identical setting except for the parameters $\omega_{i_p}$, which changed at every iteration. 

The following coupling configurations always lead to chimera states for the examined range of parameters, couplings and pinning features: $D~=~\begin{bmatrix} 0& 0 \\ 0 & 1 \end{bmatrix}~~,~~~\begin{bmatrix} 0& 1 \\ 0 & 1 \end{bmatrix}~~,~~~\begin{bmatrix} 0& 1 \\ 1 & 0 \end{bmatrix}~~,~~~\begin{bmatrix} 1& 1 \\ 0 & 0 \end{bmatrix}~~,~~~\begin{bmatrix} 0& 0 \\ 1 & 1 \end{bmatrix}~~,~~~\begin{bmatrix} 1& 1 \\ 0 & 1 \end{bmatrix} $. 

The following configurations lead $40\%-60\%$ of times to chimeras{, while the remaining times to weak chimera-like states}, depending on the parameters $\omega_{i_p}$: \\ $D~=~\begin{bmatrix} 1& 0 \\ 0 & 0 \end{bmatrix}~~,~~~\begin{bmatrix} 1& 0 \\ 0 & 1 \end{bmatrix}~~,~~~\begin{bmatrix} 1& 1 \\ 1 & 0 \end{bmatrix}~~,~~~\begin{bmatrix} 0& 1 \\ 1 & 1 \end{bmatrix} $. Despite not being as easy as in the former case, chimera states in the latter can be achieved by increasing the coupling strength, the intensity of the parameters $\omega_{i_p}$, the pinned nodes and the pervasiveness of the pinning (e.g., pinning every node instead of $1$ every $3$). 

The following coupling configurations never lead to chimera states for the examined range of parameters, couplings and pinning features: $D~=~\begin{bmatrix} 1& 0 \\ 1 & 1 \end{bmatrix}~~,~~~\begin{bmatrix} 1& 1 \\ 1 & 1 \end{bmatrix} $. 

Particularly interesting are the following coupling configurations: $D~=~\begin{bmatrix} 0& 1 \\ 0 & 0 \end{bmatrix}~~,~~~\begin{bmatrix} 0& 0 \\ 1 & 0 \end{bmatrix} $, where a chimera state emerges, but it is unstable and the incoherence region grows until the whole system develops a fully incoherent state. Moreover, such behavior is independent on the number of pinned nodes and occurs also when only one node is perturbed, which makes this setting interesting for applications in which incoherence needs to be achieved. Let us note that, in this case, higher-order interactions do not play a role, but the key feature is the coupling configuration. In fact, we obtain the same result in the pairwise setting. 

Lastly, let us point out that also different hyperring topologies and the additive pinning configuration lead to analogous behaviors and that no chimera nor incoherent states are observed when the higher-order interactions are "flattened" onto the corresponding clique-projected networks with none of the two pinning approaches. Again, let us stress that all the above does not apply to the coupling configurations leading to full incoherence, namely, $~~\begin{bmatrix} 0& 1 \\ 0 & 0 \end{bmatrix}~~,~~~\begin{bmatrix} 0& 0 \\ 1 & 0 \end{bmatrix} $, whose behavior is determined by the coupling and not by the presence of higher-order interactions. In fact, the same behavior is observed also for pairwise interactions.

\section{Scaling with respect to the number of nodes for the parametric pinning control approach}
\setcounter{equation}{0}
\renewcommand{\theequation}{B\arabic{equation}}
\setcounter{figure}{0}
\renewcommand{\thefigure}{B\arabic{figure}}\label{sec:appB}

In this Appendix, we complement the Main Text by showing the results obtained for the case of parametric pinning, which are qualitatively analogous to those of Sec. \ref{sec:numerics} obtained through the additive pinning approach, i.e., those on the scaling of the fraction of pinned nodes with respect to the hyperring size. In this setting, we are able to keep the number of pinned nodes constant as we increase the size of the structure. Again, let us stress that the total number of nodes in the hyperring increases, but the number of hyperedges is kept constant. 

\begin{figure}
\includegraphics[scale=0.3]{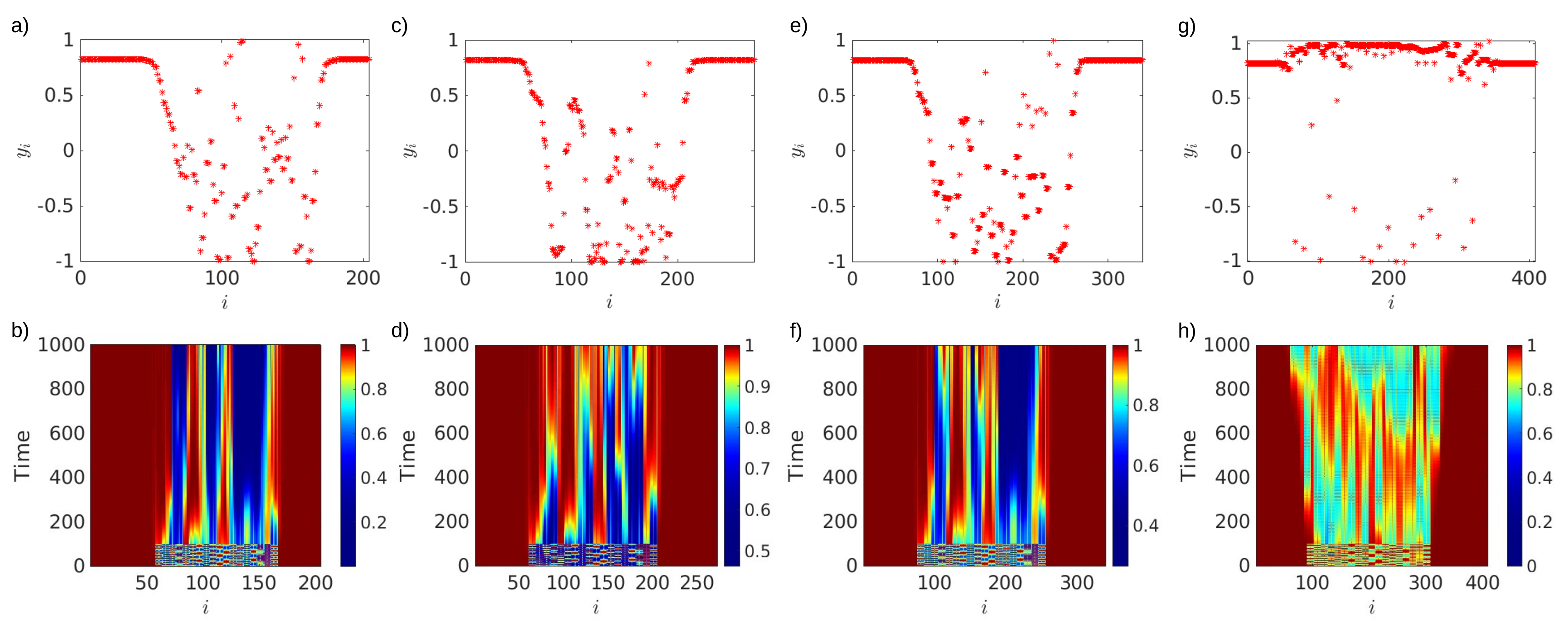}
\caption{Scaling of the pinned subset with respect to the higher-order structure size with the parametric pinning approach. On the upper panels we depict the snapshots of variables $y_i(t)$ with $i=1,...,N$ for $t_{final}=1000$ time units, while the lower panels show the hyperedge-based local order parameters. { The results for the variables $x_i(t)$ are analogous and, hence, not shown.} Panels a) and b) report the simulations performed on a $3$-hyperring (i.e., $4$-body interactions) of $204$ nodes, panels c) and d) on a $4$-hyperring (i.e., $5$-body interactions) of $272$ nodes, panels e) and f) on a $5$-hyperring (i.e., $6$-body interactions) of $340$ nodes and panels g) and h) on a $6$-hyperring (i.e., $7$-body interactions) of $408$ nodes. The number of nodes is chosen so that each hyperring is made of $68$ hyperedges. Pinning control is applied to $N_p=18$ nodes as in Fig. \ref{fig:pinning_scheme}, i.e., every two junction nodes, to all the structures and for $t_p=100$ time units. This means that the pinned nodes are $\simeq 8.8\%$ of the total nodes in the $3$-hyperring, $\simeq 6.6\%$ of the total nodes in the $4$-hyperring, $\simeq 5.3\%$ of the total nodes in the $5$-hyperring and $\simeq 4.4\%$ of the total nodes in the $6$-hyperring. The model parameters are $\alpha=1$ and $\omega=1$, the coupling strength is $\varepsilon=0.01$, except for the $6$-hyperring where $\varepsilon=0.1$, and the parameters $\omega_{i_p}$ are drawn from a uniform distribution in the interval $[1.5,5.5]$ and are the same for all the simulations.} \label{fig:scaling_par}
\end{figure}

In Fig. \ref{fig:scaling_par}, we show the results obtained with such control scheme on $d$-hyperrings with $d=3,4,5,6$, i.e., $4$-, $5$-, $6$- and $7$-body interactions, where we have fixed the number of hyperedges. Indeed, we can observe that we obtain a chimera state by inducing a large region of incoherence (more than half of the nodes) with a control that involves only a small fraction of the nodes. Moreover, the pinned nodes $N_p$ are kept constant for every structure, meaning that $N_p$ does not scale with the number of nodes, but rather with the number of hyperedges, which allows to control large structures with only a handful of nodes. The parameters $\omega_{i_p}$ are the same for all the simulations. The pinned nodes are $\simeq 8.8\%$ of the total nodes for the $3$-hyperring (panels a) and b)), $\simeq 6.6\%$ of the total nodes for the $4$-hyperring (panels c) and d)), $\simeq 5.3\%$ of the total nodes in the $5$-hyperring (panels e) and f)) and $\simeq 4.4\%$ of the total nodes in the $6$-hyperring (panels g) and h)). In the latter case, the pinning scheme consisting in controlling one every $2$ junction nodes does not yield a chimera state, {not even a weak chimera-like state,} as shown in Fig. \ref{fig:scaling_par}g-h), where, moreover, we see that the region of incoherence enlarges. This does not change if we increase the number of pinned nodes, {nor increase the coupling strength,} but only if we reduce the gap between them.

\end{document}